\documentclass[onecolumn,showpacs,fleqn,nobibnotes]{revtex4}

\usepackage{amsmath}
\usepackage{graphicx}
\usepackage{float}
\usepackage{subfigure}

\def\lsim{\raise0.3ex\hbox{$<$\kern-0.75em\raise-1.1ex\hbox{$\sim$}}}
\def\gsim{\raise0.3ex\hbox{$>$\kern-0.75em\raise-1.1ex\hbox{$\sim$}}}

\def\beq{\begin{equation}}
\def\eeq{\end{equation}}
\def\beqa{\begin{eqnarray}}
\def\eeqa{\end{eqnarray}}

\newcommand{\la}{\langle}
\newcommand{\ra}{\rangle}

\def\gappeq{\mathrel{\rlap {\raise.5ex\hbox{$>$}}
{\lower.5ex\hbox{$\sim$}}}}

\def\lappeq{\mathrel{\rlap{\raise.5ex\hbox{$<$}}
{\lower.5ex\hbox{$\sim$}}}}

\def\Toprel#1\over#2{\mathrel{\mathop{#2}\limits^{#1}}}

\begin{document}

\title{Magnetic excitation  in relativistic heavy ion collisions}
\author{I. Danhoni$^1$ and F. S. Navarra$^1$}
\affiliation{$^1$Instituto de F\'{\i}sica, Universidade de S\~{a}o Paulo, 
 Rua do Mat\~ao, 1371, CEP 05508-090,  S\~{a}o Paulo, SP, Brazil\\
}
\begin{abstract}
In this note we study  the conversion of nucleons into 
deltas induced by a strong magnetic field in ultraperipheral relativistic heavy ion 
collisions. The interaction
Hamiltonian couples the  magnetic field to the spin operator,
which, acting on the spin part of the wave function, converts a spin 1/2
into a spin 3/2 state. We estimate this transition probability and calculate the 
cross section for delta production. This process can in principle be measured, since the
delta moves close to the beam and  decays almost exclusively into pions. Forward pions may 
be detected by forward calorimeters. 
\end{abstract}
\maketitle

\section{Introduction}

In relativistic heavy ion collisions we observe extreme phenomena. One of them is the production of the strongest 
magnetic field of the universe \cite{skokov,voro,muller2}. This field is so intense because the charge density is large, 
because the speed of the source is very close to the speed of light and also because we probe it at extremely small distances 
(a few fermi) of the source. After the first estimates of $\vec{B}$,  
there has been a search for observable effects of this strong field
\cite{hattori}. The first and most famous is the Chiral Magnetic Effect (CME) \cite{cme}. Unfortunately, after ten years of 
experimental searches there are no conclusive results. From the theoretical point of view the study of the CME is quite complex.  
It is desirable to look also for conceptually simpler effects of the magnetic field, 
such as the one discussed in this note. 

We are going to study ultraperipheral relativistic heavy ion collisions (UPC's), in which the two nuclei do not overlap \cite{upc}.  
Since there is no superposition of hadronic matter, the strong interaction is strongly suppressed and the collision becomes essentially 
a very clean electromagnetic process almost without hadronic background. The few produced particles are mostly in the central rapidity region.  
In UPC's we do not expect to see produced particles at very large rapidities.  So far, the only particles measured in this region are neutrons
originated from electromagnetic excitation of the nuclei \cite{alice12}, a process in which the  energy exchanged between the nuclei is 
enough to fragment them but too small for particle production. 

We will argue that forward pions are very likely to be produced by magnetic excitation (ME) of the nucleons in the nuclei. The strong magnetic 
field produced by one nucleus induces magnetic transitions, such as $N \to \Delta$ (where $N$ is a proton or a neutron), in the 
nucleons of the other nucleus. The produced $\Delta$ keeps moving together with the nucleus (or very close to it) and then decays almost 
exclusively through the reaction $\Delta \to N + \pi$. From the kinematics of this two-body decay we know that the pion has momentum around 
$\simeq 200$ MeV in the $\Delta$ rest frame. Thus, in the cms frame it has a very large longitudinal momentum and very large rapidity. Since
there is no other competing mechanism for forward pion production, the observation of these pions would be a signature of the magnetic 
excitation of the nucleons and also an indirect measurement of the magnetic field.  In what follows we will show that ME has a very large cross section.

\section{Formalism}

Magnetic excitation in the context of heavy ion collisions was first considered 
in \cite{muller1}, where it was argued that in the presence of a strong magnetic 
field the transition $\eta_c \to J/\psi$ might happen. Unfortunately, the transition 
rate was extremely small. This was due to the presence of the charm quark mass in the 
denominator of the transition amplitude. Here we revisit the idea, using the same formalism and applying it to 
the nucleon-delta transitions in ultraperipheral collisions. Now, instead of the 
heavy quark mass we have a light quark mass in the denominator and, as will be 
seen,  the transition rate becomes large.

Under the influence of a strong magnetic field, a nucleon is converted into a $\Delta$. For the sake of definiteness let us consider the 
transition $|n \uparrow \rangle  \to | \Delta^0 \uparrow \rangle$. 
The amplitude for this process is given by:
\begin{equation}
    a_{fi} = -i \int_{-\infty}^{\infty} e^{iE_{fi}t'}
\langle \Delta^0\uparrow| H_{int} (t') |n \uparrow \rangle \, dt'
\label{amp} 
\end{equation}
where $\hbar = 1$ and where
\begin{equation}
E_{fi}=\frac{m_\Delta^2-m_n^2}{2m_n}
\end{equation}
In the above equation $m_{\Delta}$ and $m_p$ are the $\Delta$ and nucleon masses
respectively. The interaction Hamiltonian is given by:
\begin{equation}
    H_{int}= - \vec \mu . \vec B
\end{equation}
The magnetic dipole moment of the nucleon is given by the sum of the  magnetic dipole moments of the corresponding constituent quarks:
\begin{equation}
    \vec \mu = \sum_{i=u,d}\vec \mu_i = \sum_{i=u,d} \frac{q_i}{m_i}\vec S_i
\end{equation}
where $q_i$ and $m_i$ are the charge and mass of the  quark of type $i$ and  $\vec{S}_i$ is the spin operator acting on the spin state of  this quark. 
\begin{figure*}[t]
\includegraphics[scale=0.30]{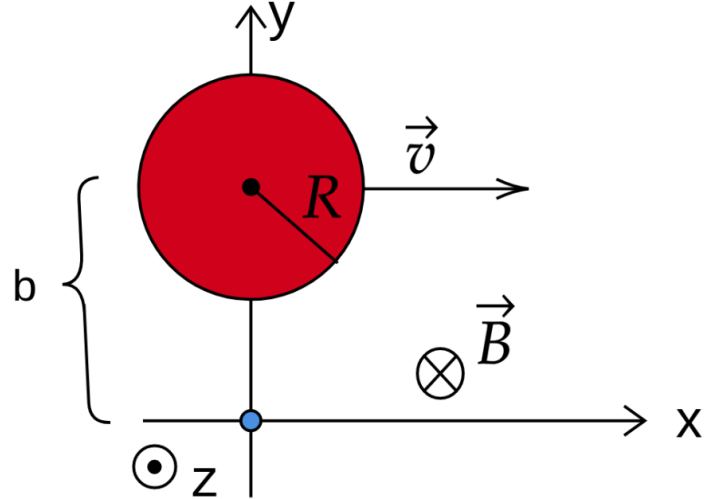} 
\caption{Coordinate system with magnetic field along the z direction. 
The projectile  nucleus  moving  with velocity $\vec{v}$    
at impact parameter $b $. The blue circle represents a 
nucleon at rest.}
\label{fig1}
\end{figure*}
In Fig. \ref{fig1} we show the system of coordinates and the moving projectile. The projectile moves along the x direction 
and the  magnetic field is in  the z direction.  Because of the symmetry of the problem  
there is  no magnetic field in the collision plane. Since we are studying an UPC, 
we will, for simplicity, assume that the projectile-generated field is the
same produced by a point charge. The field is given by \cite{muller2}:
\begin{equation}
   B_z= \frac{1}{4\pi}\frac{qv\gamma (b-y)}{((\gamma(x-vt))^2
        +(y-b)^2+z^2)^{3/2}}
\label{field}
\end{equation}
In the above expression $\gamma$ is the Lorentz factor, $b$ is the impact parameter along the $y$ direction, $v \simeq 1$ is the 
projectile velocity and the projectile electric charge is $q =Z e$.

The interaction Hamiltonian acts on spin states. The relevant ones are:
\begin{equation}
|p    \uparrow \ra=\frac{1}{3\sqrt{2}}[udu(\downarrow\uparrow\uparrow+\uparrow\uparrow\downarrow  
-2\uparrow\downarrow\uparrow)+duu(\uparrow\downarrow\uparrow+\uparrow\uparrow\downarrow
-2\downarrow\uparrow\uparrow)+uud(\uparrow\downarrow\uparrow+\downarrow\uparrow\uparrow-2\uparrow\uparrow\downarrow)]
\end{equation}
\begin{equation}
| p \downarrow \ra=\frac{1}{3\sqrt{2}}
[udu(\uparrow\downarrow\downarrow + \downarrow\downarrow\uparrow
-2\downarrow\uparrow\downarrow)+duu(\downarrow\uparrow\downarrow
+\downarrow\downarrow\uparrow-2\uparrow\downarrow\downarrow)
+uud(\downarrow\uparrow\downarrow 
+\uparrow\downarrow\downarrow-2\downarrow\downarrow\uparrow)]
\end{equation}
\begin{equation}
|n   \uparrow \ra=\frac{1}{3\sqrt{2}}[dud(\downarrow\uparrow\uparrow 
+\uparrow\uparrow\downarrow-2\uparrow\downarrow\uparrow)
+udd(\uparrow\downarrow\uparrow+\uparrow\uparrow\downarrow
-2\downarrow\uparrow\uparrow)+ddu(\uparrow\downarrow\uparrow
+\downarrow\uparrow\uparrow-2\uparrow\uparrow\downarrow)]
\end{equation}
\begin{equation}
|n \downarrow \ra=\frac{1}{3\sqrt{2}}[dud(\uparrow\downarrow\downarrow 
+\downarrow\downarrow\uparrow-2\downarrow\uparrow\downarrow)
+udd(\downarrow\uparrow\downarrow +\downarrow\downarrow\uparrow
-2\uparrow\downarrow\downarrow)+ddu(\downarrow\uparrow\downarrow
+\uparrow\downarrow\downarrow-2\downarrow\downarrow\uparrow)]
\end{equation}
\begin{equation}
|\Delta^+ \uparrow\ra=\frac{1}{3}(uud+udu+duu)(\uparrow\uparrow\downarrow
+\uparrow\downarrow\uparrow+\downarrow\uparrow\uparrow)
\end{equation}
\begin{equation}
|\Delta^+ \downarrow\ra=\frac{1}{3}(uud+udu+duu)(\downarrow\downarrow\uparrow
+\downarrow\uparrow\downarrow+\uparrow\downarrow\downarrow)
\end{equation}
\begin{equation}
|\Delta^0 \uparrow\ra=\frac{1}{3}(ddu+dud+udd)(\uparrow\uparrow\downarrow
+\uparrow\downarrow\uparrow+\downarrow\uparrow\uparrow)
\end{equation}
\begin{equation}
|\Delta^0 \downarrow\ra=\frac{1}{3}(ddu+dud+udd)(\downarrow\downarrow\uparrow
+\downarrow\uparrow\downarrow+\uparrow\downarrow\downarrow)
\label{estadospin}
\end{equation}
With these ingredients we can compute all the matrix elements:
$\la\Delta^+\uparrow|H_{int}|p \uparrow\ra$,
$\la\Delta^+ \downarrow|H_{int}|p \downarrow\ra$,
$\la\Delta^0 \uparrow|H_{int}|n \uparrow \ra$ and 
$\la\Delta^0 \downarrow|H_{int}|n \downarrow\ra$.
The required matrix elements can be obtained by substituting Eq.  (4) into Eq. (3) and then calculating the sandwiches
of Eq. (3) with the spin states given above. For example, 
in the case of $\la\Delta^0 \uparrow|H_{int}|n \uparrow\ra$ we have: 
\begin{equation}
\la \Delta^0\uparrow|H_{int}|n\uparrow\ra
=-\frac{B_z}{m}\sum_{q=u,d} \la\Delta^0\uparrow|e_q S_{qz}|n\uparrow\ra
\end{equation}
where we have assumed that all the light quarks have the same constituent quark mass
$m$. 
As the charges of the quarks up and down are known to be $e_u=2e/3$ e $e_d=-e/3$, applying the $S_z$ operator  on the neutron 
wavefunction yields:
\begin{eqnarray}
\nonumber
   e_qS_{qz} |n \uparrow\ra & = & \frac{1}{6\sqrt{2}}\Bigg[\frac{2e}{3} |d^\downarrow u^\uparrow d^\uparrow\ra+ 
\frac{2e}{3}|d^\uparrow u^\uparrow d^\downarrow\ra+\frac{4e}{3}|d^\uparrow u^\downarrow d^\uparrow\ra+
\frac{2e}{3}|u^\uparrow d^\downarrow d^\uparrow\ra+\frac{2e}{3}|u^\uparrow d^\uparrow d^\downarrow\ra+ \\
    &  & \frac{8e}{3}|u^\downarrow d^\uparrow d^\uparrow\ra+\frac{2e}{3} |d^\uparrow d^\downarrow u^\uparrow\ra+ 
\frac{2e}{3}|d^\downarrow d^\uparrow u^\uparrow\ra+\frac{8e}{3}|d^\uparrow d^\uparrow u^\downarrow\ra\Bigg]
\end{eqnarray}
Multiplying the above expression by  the $\Delta$ wavefunction we obtain:
\begin{equation}
\la\Delta^0\uparrow|H_{int}|n \uparrow\ra=-\frac{\sqrt{2}Be}{3m}
\end{equation}
Evaluating all the possible nucleon-delta transition matrix elements we find:  
$$
\la\Delta^0\uparrow|H_{int}|n \uparrow\ra \, = \, \la \Delta^+ \downarrow|H_{int}|p \downarrow\ra \, 
= \, -\frac{\sqrt{2}Be}{3m}
$$
$$
\la \Delta^0 \downarrow|H_{int}|n \downarrow\ra \, 
= \, \la \Delta^+\uparrow|H_{int}|p \uparrow\ra \, =\, \frac{\sqrt{2}Be}{3m}
$$
The cross section for a single $N \to \Delta$ transition is given by:
\begin{equation}
    \sigma=\int |a_{fi}|^2 \, d^2b = 2\pi\, \int|a_{fi}|^2\, b \, db 
\end{equation}
where we have used cylindrical symmetry  $d^2b\, =\, b\, db\, d\theta \to 2\, \pi\,  b \, db$. Inserting the above matrix element 
into (\ref{amp}) and using it in the above expression we find:
\begin{equation}
\sigma    =\frac{Z^2e^4}{9\pi m^2} \left( \frac{E_{fi}}{v\gamma} \right)^2 \int_{R}^{\infty}
\Big[ K_1\Big(\frac{E_{fi}b}{v\gamma}\Big)\Big]^2 b \, db
\label{crossmid}
\end{equation}
where $K_1$ is the modified Bessel function and $\gamma = \sqrt{s}/ (2 m_n)$. 
When the target is a nucleus all nucleons  have the same squared transition amplitude. Hence the nuclear target 
provides a nucleon flux enhancing  the cross section which is, in a first approximation, given by:
\begin{equation}
\sigma_A \, \simeq  \, A \, \sigma
\label{crossfin}
\end{equation}
Before presenting the numerical evaluation of (\ref{crossmid}) and (\ref{crossfin}) 
it is useful to make some analytical estimates. 
The typical strength of the magnetic fields can be roughly estimated from (\ref{field}) 
(or taken from \cite{hattori}) and is given by:
\beq
eB \simeq \frac{\gamma \alpha_{em} Z}{R^2_A}  
\eeq
where $R_A$ is the nuclear radius. The matrix elements can be approximated as:
\beq
\la \Delta | H_{int}| n \ra = \la \Delta | \vec{\mu} . \vec{B}| n \ra = 
\la \Delta | \frac{e B S_z}{m}| n \ra \simeq \frac{\gamma \alpha_{em} Z}{m R^2_A}
\eeq
Inserting the above expression into (\ref{amp}), assuming that $E_{fi}$ is small and 
that $H_{int}(t')$ is different from zero only during the collision, i.e., in the time interval $\Delta t \simeq R_A$, 
the transition amplitude  simplifies to:
\beq
a_{fi} \simeq \la \Delta | \frac{e B S_z}{m}| n \ra \, R_A =
\frac{\gamma \alpha_{em} Z}{m R_A} 
\eeq
Finally the single $N \to \Delta$  transition cross section  reads:
\beq
\sigma \simeq  |a_{fi}|^2  \,\,   R^2_A 
\simeq \left( \frac{\gamma \alpha_{em} Z}{m R_A} \right)^2    R^2_A
\simeq \frac{ \gamma^2 \alpha_{em}^2 Z^2}{m^2}  
\label{fcross}
\eeq
\section{Results and discussion}

\begin{figure*}[t]
\includegraphics[scale=0.70]{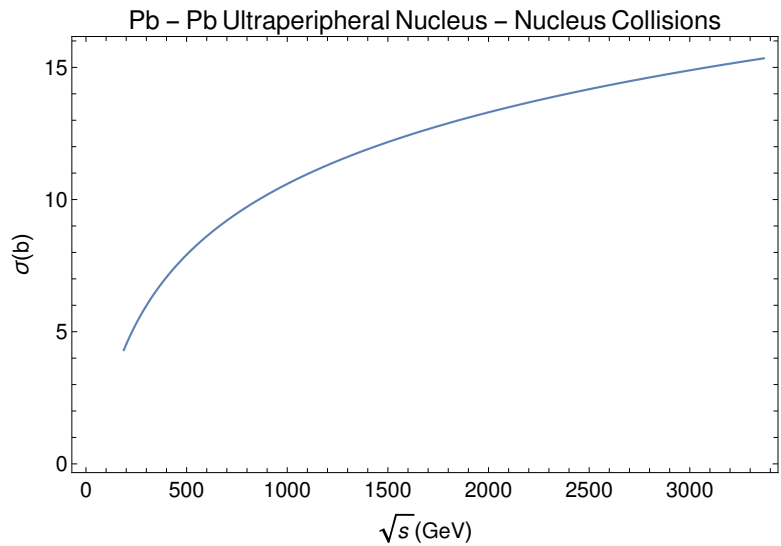}
 \caption{Magnetic excitation cross section.}
\label{fig2}
\end{figure*}

Let us consider a lead ($Z=82$) - proton  collision. We will  use 
$m=0.36$ GeV, $R=7$ fm and $v=1$. Integrating (\ref{crossmid}) and plotting it 
as a function of $\sqrt{s}$ we obtain the result shown in  Fig. \ref{fig2}. 
The cross section obtained in our calculations  illustrates the enormous effect that  the 
magnetic field can produce in ultraperipheral collisions. The ``pocket formula'' (\ref{fcross}) gives the 
correct order of the magnitude.

Whenever the nucleon excitation to a delta resonance occurs, this latter will decay with 99 \% probability into a 
nucleon and a pion. Hence, the above number is also the cross section for pion production through $\Delta$ decay. 
Since the magnetic field does not transfer any momentum to the nucleon, when it is converted to a $\Delta$ the resonance keeps 
moving together with the nuclear 
target. 
When the $\Delta$ decays, the outcoming pion has a momentum of $p_{\pi} \simeq 200$ MeV in the target restframe. Hence the pion 
will escape from the nucleus but
will be moving in the same forward rapidity region. 
Our result indicates that forward pion production has a large cross section and could be observed by the ALICE collaboration  
in a similar way it was done for neutrons in Ref. \cite{alice12}. 

Our calculation can be improved. In particular we should use the correct spatial charge distributions instead of the pointlike charge 
approximation. Moreover we should include recoil effects due to the electric field. Finally we should  compute the pion 
rapidity and transverse momentum distributions. This would be crucial to study the possibility of detecting them. 
Work along this direction is already in progress.


\begin{acknowledgments}
The authors thank G. Germano, M. Cosentino, M. Munhoz and B. Muller for helpful comments. 
This work was  partially financed by the Brazilian funding agencies CAPES and CNPq.
\end{acknowledgments}


\end{document}